\documentclass[usenatbib]{mnras}  

\pdfoutput=1

\usepackage{newtxtext,newtxmath}
\usepackage[T1]{fontenc}
\usepackage{ae,aecompl}

\usepackage{amsmath} 
\usepackage{graphicx}

\hypersetup{breaklinks,
            pdftitle={The Voigt and complex error function: Humlicek's rational approximation revisited},
            pdfkeywords={},
            pdfkeywords={Complex probability function; Plasma dispersion function; Fadde(y)eva function},
            pdfauthor={F. Schreier}}

\batchmode

\newcommand{\D}{\mathrm{d}}
\newcommand{\E}{\mathrm{e}}
\newcommand{\I}{\mathrm{i}}
\newcommand{\chem}[1] {{\ensuremath{\mathrm{#1}}}}
\newcommand{\qeq}[1]  {Eq.~(\ref{#1})}
\newcommand{\qufig}[1] {Fig.~\ref{#1}}
\newcommand{\qutab}[1] {Table~\ref{#1}}
\newcommand{\ten}[1] {\ensuremath{\cdot 10^{#1}}}

\title[Huml\'\i\v{c}ek's complex error function algorithm]
      {The Voigt and complex error function: \\
      Huml\'\i\v{c}ek's rational approximation generalized}

\author[F. Schreier]{Franz Schreier \thanks{E-mail: franz.schreier@dlr.de} \\
        DLR --- Deutsches Zentrum f\"ur Luft- und Raumfahrt, 
        Institut f\"ur Methodik der Fernerkundung, 
        82234 Oberpfaffenhofen, Germany}

\date{Accepted 2018 June 20. Received 2018 June 18; in original form 2018 May 2}
\pubyear{2018}

\begin{document}

\label{firstpage}
\pagerange{\pageref{firstpage}--\pageref{lastpage}}
\maketitle

\begin{abstract}
Accurate yet efficient computation of the Voigt and complex error function is a challenge since decades in astrophysics and other areas of physics.
Rational approximations have attracted considerable attention and are used in many codes, often in combination with other techniques.
The 12-term code ``\texttt{cpf12}'' of Huml\'\i\v{c}ek (1979) achieves an accuracy of five to six significant digits throughout the entire complex plane.
Here we generalize this algorithm to a larger (even) number of terms.
The $n=16$ approximation has a relative accuracy better than $10^{-5}$ for almost the entire complex plane except for very small imaginary values of the argument even without the correction term required for the \texttt{cpf12} algorithm.
With 20 terms the accuracy is better than $10^{-6}$.
In addition to the accuracy assessment we discuss methods for optimization and propose a combination of the 16-term approximation with the asymptotic approximation of Huml\'\i\v{c}ek (1982) for high efficiency.
\end{abstract}



\section{Introduction}
\label{sec:introduction}

The Voigt profile \citep[][print companion to the {NIST} Digital Library of Mathematical Functions, \url{http://dlmf.nist.gov/}]{NIST-HMF}, i.e.\ the convolution of a Lorentzian and Gaussian profile, is ubiquitous in many branches of physics including astrophysics, e.g., \citet{Peraiah02,Heng17b}. 
From a computational point it is more convenient to consider the Voigt function $K(x,y)$ depending on two variables $x$ and $y$
(essentially the distance from the center peak and the ratio of the Lorentz to Gauss width)
and the complex error function $w(x+\I y)$, whose real part gives the Voigt function.
Because there is no closed-form solution for the convolution integral, numerous algorithms have been developed in the past for efficient and/or accurate evaluation
utilizing series or asymptotic expansions, continued fractions, rational approximations, Gauss-Hermite quadrature etc.
\citep[for an old review, still worth while to read, see][]{Armstrong67}.

For an assessment of Voigt and complex error function algorithms the intended application has to be considered.
Highly accurate codes such as \citet{Poppe90,Poppe90a} (14 significant digits stated accuracy) and \citet{Zaghloul11} (20 significant digits) or arbitrary precision codes \citep{Boyer14,Molin11} are indispensable for tests of an algorithm's accuracy, but not necessarily fast.
High resolution line-by-line (lbl) radiative transfer modeling \citep{Bailey12,Schreier14} presents a ``million- to billion-line challenge'' \citep{Grimm15,Heng17b} where numerous Voigt functions have to be evaluated \citep{Rothman10,Tennyson12} at thousands to millions of frequency grid points and computational speed becomes a prime concern.

In view of the few digits used for line strengths and broadening parameters in spectroscopic data bases such as HITRAN \citep{Gordon17}, HITEMP \citep{Rothman10}, or GEISA \citep{JacquinetHusson16}, four significant digits for the Voigt function can be regarded as appropriate for many lbl calculations \citep{Schreier11v}.
On the other hand, the increasing quality of spectroscopic measurements has indicated deficiencies of the Voigt profile.
More sophisticated line shapes accounting for collisional Dicke narrowing, line mixing, or speed-dependence are required \citep{Varghese84,Tennyson14}
that can often be expressed in terms of the complex error function, where the imaginary parts of the arguments can differ by more than ten orders of magnitude \citep{Tran13,Schreier17}.
For ``standard'' Voigt lbl modeling in astrophysical and planetary atmospheric spectroscopy $y$ can be as large as $10^5$ and as small as $10^{-8}$
\citep{Lynas-Gray93,Wells99,TepperGarcia06,TepperGarcia07e,Schreier11v}.

Rational approximations are particularly attractive because they can be implemented efficiently (i.e.\ only one division plus additions and multiplications) and allow for high accuracy.
\citet{Hui78} presented a single rational approximation with a fifth-degree numerator polynomial and a sixth-degree denominator polynomial that should be ``sufficiently accurate for most applications over the whole complex plane''.
\citet{Humlicek79} proposed a 12-term rational approximation ``\texttt{cpf12}''; including a modification for small $y$ the maximum relative error is less than $5 \cdot 10^{-6}$.
\citet{Humlicek82} suggested a division of the complex plane into four subdomains; using appropriate rational approximations the ``\texttt{w4}'' code evaluates the complex error function accurate to four significant digits. 
\citet{Weideman94} developed a single rational approximation applicable in the entire complex plane whose accuracy can be adjusted by selection of the number $N$ of terms.
\citet{Kochanov11v} presented two sums with four and six fractions with complex coefficients.

Problems of the \citet{Hui78} code for small $y$ have been reported early by \citet{Karp78} and \citet{Humlicek82} \citep[see also][Fig.\ 4]{Schreier11v}.
The \citet{Kochanov11v} approximation shows significant errors for small $y < 10^{-2}$.
\citet{Shippony93,Shippony03} use the \citeauthor{Hui78} approach (for small $|x|$ and intermediate $y$) along with other techniques.
\citeauthor{Humlicek79}'s codes appear to belong to the most popular complex error function codes
(according to Scopus (March 2018), there are 218 and 313 citing articles for the 1979 and 1982 paper, respectively).
In particular the \texttt{w4} code has been further developed by \citet{Kuntz97,Ruyten04,Imai10}
and the \texttt{cpf12} code has been implemented by \citet{Tran13} for evaluation of the ``Hartmann-Tran'' profile \citep{Tennyson14}.
\citet{Wells99} combined approximations of both \citeauthor{Humlicek79} codes with slightly modified region bounds.
\citet{Zaghloul17} allows to select the accuracy and uses \citet{Humlicek82} for low accuracy.

Except for \citet{Weideman94} all codes achieve moderate to high accuracy throughout the whole complex plane only by a combination of several methods.
Unfortunately this can make the code difficult to optimize; this is especially true if the conditional branches are more than a simple ``either-or'' as for example in the \citet{Humlicek79} \texttt{cpf12} algorithm.
To avoid complicated (e.g.\ nested) \texttt{if} structures we have combined the asymptotic rational approximation of \citet{Humlicek82} for $|x|+y>15$ with the \citet{Weideman94} approximation \citep{Schreier11v}.

Whereas the four rational approximations of \citet{Humlicek82} have been utilized by several refinements, the somewhat older and slightly more accurate \citet{Humlicek79} \texttt{cpf12} code has rarely been used in other complex error function codes (except for \citet{Wells99}).
In particular we are not aware of any algorithm using the \citet{Humlicek79} rational approximation with a higher (or smaller) number of terms (i.e.\ $n \ne 12$).

The objective of this paper is an assessment of the \citet{Humlicek79} rational approximation for an ``arbitrary'' number of terms.
We briefly review some basic facts in the following section, followed by a presentation of the \citeauthor{Humlicek79}  algorithm, and a discussion of our optimization strategies.
In Section \ref{sec:results} we evaluate the accuracy and efficiency of the ``generalized'' \citet{Humlicek79} code, and provide our conclusions in Section \ref{sec:conclusions}.

\section{Theory and Methods}
\label{sec:theory}

\subsection{The Voigt and complex error function}
\label{ssec:voigt}

The Voigt (or Hjerting) profile and the closely related Voigt function are defined by
\begin{align}
 g_\text{V}(\nu-\hat\nu,\gamma_\text{L},\gamma_\text{G})
      ~&=~ \int_{-\infty}^\infty \D\nu' \; g_\text{L}(\nu-\nu',\gamma_\text{L}) \,\times\, g_\text{G}(\nu'-\hat\nu,\gamma_\text{G})  \label{vgtProfile} \\
       &=~ {\sqrt{\ln 2 / \pi} \over \gamma_\text{G}} ~K(x,y) \notag \\
  K(x,y) ~&=~ {y \over \pi} ~ \int_{-\infty}^\infty {\E ^{-t^2} \over (x-t)^2 + y^2} ~ \D t ~, \label{vgtFct}
\end{align}
where $\gamma_\text{L}$ and $\gamma_\text{G}$ are the half widths at half maximum (HWHM) of the Lorentzian $g_\text{L}$ and Gaussian $g_\text{G}$, respectively,
and $\hat\nu$ is the center wavenumber or frequency (i.e.\ corresponding to the energy difference of an atomic or molecular transition).
The dimensionless arguments of the Voigt function are defined as ratios
\begin{equation}\label{defxy}
 x ~=~ \sqrt{\ln 2}~ {\nu - \hat\nu \over \gamma_\text{G}} 
\qquad\hbox{ and }\qquad
 y ~=~ \sqrt{\ln 2}~ {\gamma_\text{L} \over \gamma_\text{G}} ~.
\end{equation}
Note that the profiles in \qeq{vgtProfile} are normalized to one, $\int g(\nu) \, \D \nu = 1$, whereas the Voigt function is normalized to $\sqrt\pi$.

The Voigt function is closely related to the complex error function (a.k.a.\ complex probability function, Fadde(y)eva function, or plasma dispersion function)
\begin{equation}\label{wDef}
 w(z) ~\equiv~ K(x,y) \,+\, \I L(x,y) ~=~ {\I \over \pi} ~ \int_{-\infty}^\infty ~ {e^{-t^2} \over z-t} ~ \D t
\end{equation}
with  $z = x + \I y$.
Asymptotically the complex error function behaves as $\I / (\sqrt\pi z)$, corresponding to a Lorentz-like Voigt function for large $x$.
For $y=0$ the real part becomes a Gauss function, i.e.\ $K(x,0) = \text{Re}\bigl(w(x)\bigr) = \exp(-x^2)$.
The half width at half maximum of the Voigt function can be estimated as  \citep{Olivero77,Schreier11v}
\begin{equation} \label{xHalf}
 x_{1/2} ~=~  \dfrac{1}{2} \: \left( y \,+\, \sqrt{ y^2 + 4 \ln 2} \right) ~.
\end{equation}

\subsection{The rational approximation by \citet{Humlicek79}}
\label{ssec:humlicek}

\citeauthor{Humlicek79} presented a rational approximation in the form
\begin{equation} \label{cpf12a}
 w(z) ~=~ \sum_{\substack{k=-n/2 \\ k\ne 0}}^{n/2}  {\alpha_k + \I \beta_k \over z - x_k +\I \delta }  \\
      ~=~ \sum_{k=1}^{n/2} \left( {\alpha_k + \I \beta_k \over z - x_k +\I \delta }
                             ~-~  {\alpha_k - \I \beta_k \over z + x_k +\I \delta } \right)
\end{equation}
with real constants
\begin{align}
 \alpha_k ~&=~ -\alpha_{-k} ~=~ - {1 \over \pi} \omega_k \E^{\delta^2} \sin(2 x_k \delta) \\
 \beta_k  ~&=~ +\beta_{-k}  ~=~ + {1 \over \pi} \omega_k \E^{\delta^2} \cos(2 x_k \delta) ~. 
\end{align}
Here $x_k$ and $\omega_k$ are the roots and weights of the $n$-point Gauss-Hermite formula (with $n$ even and the nodes and weights antisymmetric and symmetric in $k$ \citep[e.g.][Table 25.10]{AbSt64}),
and $\delta$ is a positive constant to be chosen appropriately.
Note that the constants $\alpha_k$, $\beta_k$, $x_k$, $\omega_k$, and $\delta$ depend on the number $n$ of nodes,
but the superscript $(n)$ used in the original paper has been omitted here; furthermore, \citeauthor{Humlicek79} uses the notation $\delta_k^{(n)} \equiv y_0^{(n)}$ synonymously.
As emphasized by \citeauthor{Humlicek79}, ``the choice of $n$ and $\delta$ involves a compromise between several demands'',
i.e.\ a larger number of terms increases the quality of the approximation, but also the computational effort.
Table 1 of \citet{Humlicek79} gives some hints on the proper choice of $\delta$ for $8 \le n \le 20$. 

To circumvent accuracy problems near the real axis where \eqref{cpf12a} fails to approximate an almost Gaussian function (see subsection \ref{ssec:acc}), \citeauthor{Humlicek79} introduced a modification for the real part
\begin{align} \label{cpf12b}
 K(x,y) ~=~ \E^{-x^2} + & \sum_{\substack{k=-n/2 \\ k\ne 0}}^{n/2}  {y \over (x-x_k)^2 + \delta^2} \,\times \\
            & {\beta_k [ (x-x_k)^2 - \delta (y+\delta)] - \alpha_k(x-x_k)(y+2\delta) \over (x-x_k)^2 + (y+\delta)^2} \notag
\end{align}
for $y < 0.85$ and $|x| > 18.1 y+1.65$.
The ``\texttt{cpf12}'' code with $n=12$ and $\delta=1.5$ achieves a maximum relative error less than $2 \cdot 10^{-6}$ and $5 \cdot 10^{-6}$ for the real and imaginary part, respectively.

\subsection{Optimization and a combination with the \citet{Humlicek82} asymptotic approximation}
\label{ssec:combine}

Equations \eqref{cpf12a} and \eqref{cpf12b} are suboptimal w.r.t.\ computational efficiency because of the numerous divisions:
According to \citet{Ueberhuber97} divisions are significantly slower than additions or multiplications, and
\citet{Goedecker01} noted that ``the calculation of special functions, such as divisions, square roots, exponentials and logarithms requires anywhere from a few dozen cycles up to hundreds of cycles \dots\
these calculations have to be decomposed into a sequence of elementary instructions such as multiplies and adds.''
However, using elementary calculus \eqref{cpf12a} can be rewritten to the ``standard form'' of a rational approximation 
\begin{equation} \label{RatFunComplex}
w(z) ~=~ {P(z) \over Q(z)} ~=~ {\sum\limits_{k=0}^{n-1} a_k z^k \over \sum\limits_{l=0}^n b_l z^l}
\end{equation}
Note that $a_{n-1}=\I/\sqrt\pi$ and $b_n=1$ in accordance with the asymptotic expansion of the complex error function.
The numerator coefficients $a_k$ are real for even $k$ and purely imaginary for odd $k$; furthermore $b_l=0$ for all odd $l$.
Finally it is worth to mention that the coefficients depend on the parameter $\delta$.


For the 16-term approximation transformed into the form \eqref{RatFunComplex} the number of adds and multiplies is similar to the $N=32$ rational approximation of \citet{Weideman94}.
However, the benchmark tests in \citet{Schreier11v} have indicated that this approximation is significantly slower than the asymptotic rational approximation 
\begin{equation} \label{w4asymp}
 R_{1,2}(z) = {\I z /\sqrt{\pi} \over  z^2 - {1\over 2}} 
\end{equation}
used by \citet{Humlicek82} for $s=|x|+y>15$.
(The subscript on the lefthand side indicates the degree of the numerator and denominator polynomial.)
Note that for lbl calculations function values for only a few grid points have to be evaluated in the line center, and most values have to be evaluated in the line wings with large $|x|$.
Accordingly we had suggested to use the \citeauthor{Weideman94} approximation ($N=24$) near the line center only and approximation \eqref{w4asymp} otherwise.
Similarly we propose a combination of the two \citet{Humlicek79,Humlicek82} approximations
\begin{equation} \label{hum1cpf16}
 w(z) ~=~
 \begin{cases} 
   {\I z/\sqrt{\pi} \over z^2- {1 \over 2}} & |x|+y>15 \\
   {\sum\limits_{k=0}^{n-1} a_k z^k \over \sum\limits_{l=0}^n b_l z^l} \qquad  & \text{otherwise.} 
 \end{cases} 
\end{equation}


\section{Results}
\label{sec:results}

In the first subsection we provide an assessment of the accuracy of the \citeauthor{Humlicek79} rational approximation \eqref{cpf12a} for $n \ge 12$.
As an accuracy reference SciPy's \texttt{wofz} function is used,
a combination of the \citet{Poppe90,Poppe90a} and \citet{Zaghloul11} codes with a stated accuracy of at least 13 significant digits
(Scientific Python \texttt{scipy.special.wofz} implementation based on S.G. Johnson's ``Faddeeva package'', \url{http://ab-initio.mit.edu/Faddeeva}).
Preliminary timing tests are presented in Subsection \ref{ssec:ipy}, and the performance of lbl cross section modeling is discussed in the third subsection.
The combination of \eqref{cpf12a} and \eqref{cpf12b} with $n=12$ is denoted as \texttt{cpf12}, whereas the optimized form \eqref{RatFunComplex} with $n=12$ or $n=16$ is denoted as \texttt{zpf12} and \texttt{zpf16}, respectively.

\begin{figure*}
 \includegraphics[width=\textwidth]{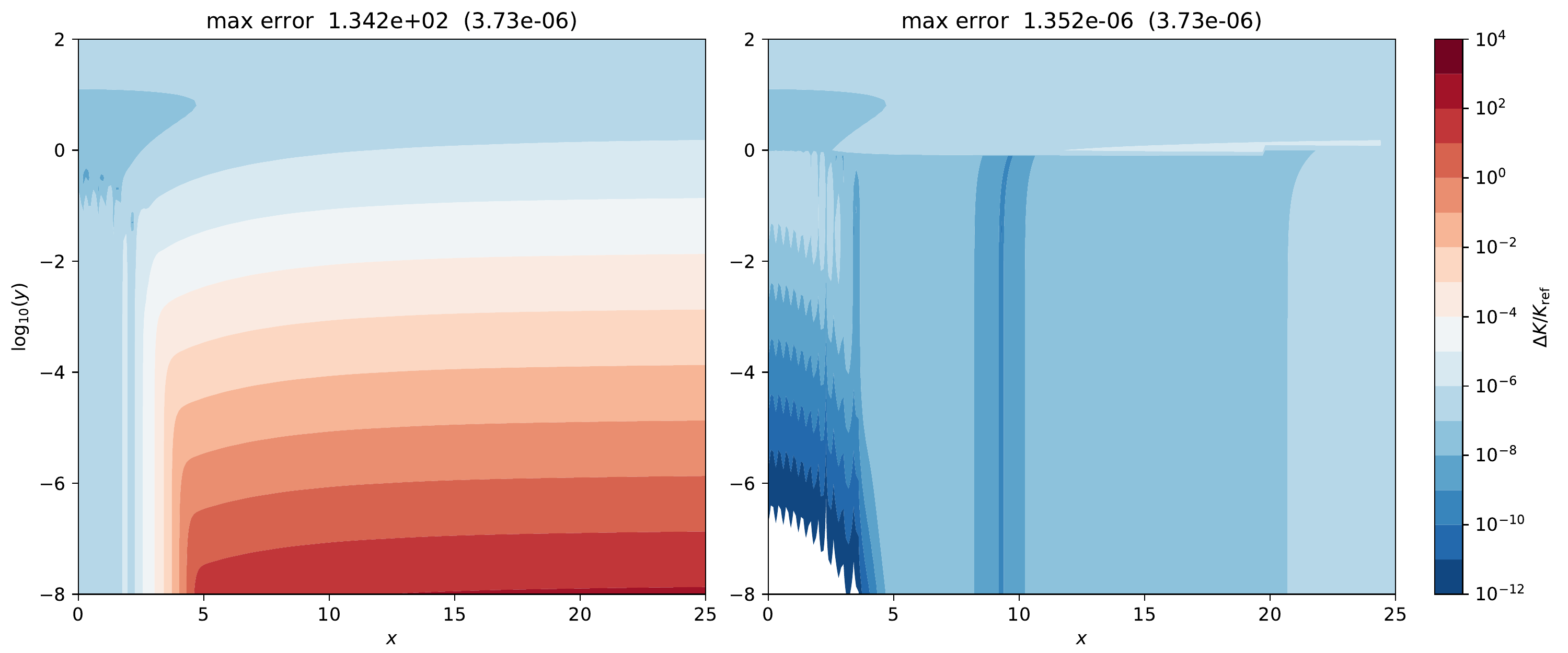}
 \caption{Comparison of the \citeauthor{Humlicek79} \texttt{cpf12} approximation for $\delta=1.5$ with the \texttt{wofz} reference.
          Left: approximation \eqref{cpf12a};  Right: \eqref{cpf12a} and \eqref{cpf12b} combined.
          White areas indicate a relative error better than $10^{-12}$.
          The numbers in the title indicate the maximum relative error of $K$ and $L$ (in parentheses).}
 \label{fctError12}
\end{figure*}

\subsection{Accuracy} \label{ssec:acc}

\qufig{fctError12} depicts the relative error $\epsilon(x,y) = |K_\text{cpf} - K_\text{wofz}| / K_\text{wofz}$ of the real part of the $n=12$ approximation.
\qufig{fctError12}a clearly confirms \citeauthor{Humlicek79}'s statement w.r.t.\ accuracy problems of the approximation \eqref{cpf12a} for small $y$
(the imaginary part $L(x,y)$ is computed correctly with a relative accuracy better than $4 \cdot 10^{-6}$).
Including the correction term \eqref{cpf12b} drastically reduces the errors, and the maximum relative error of about $ 10^{-6}$ is in accordance with \citeauthor{Humlicek79}'s claims.
However, the drawback of the correction term is the difficulty to implement this efficiently.

\begin{figure*}
 \includegraphics[width=\textwidth]{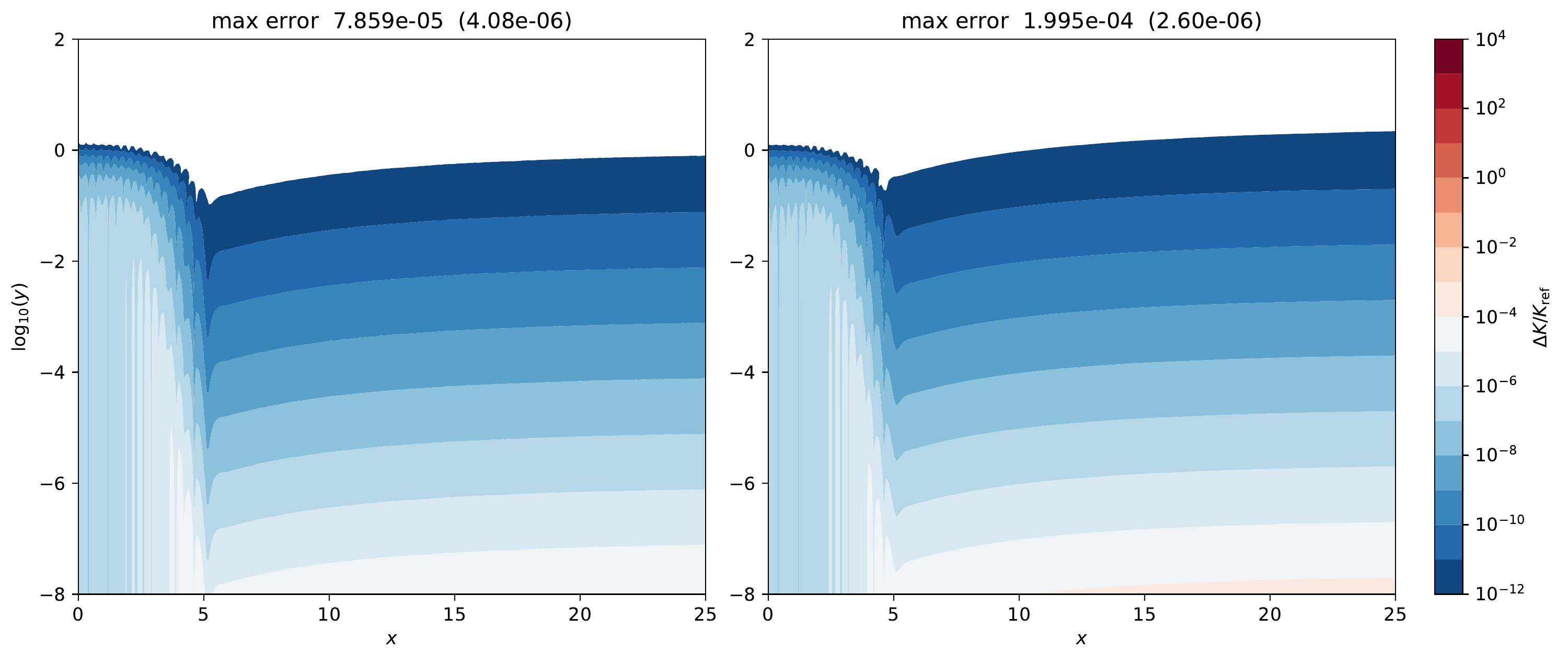}
 \caption{Comparison of \citeauthor{Humlicek79} \texttt{zpf16} approximation for $\delta=1.3118$ (left) and $\delta=1.35$ (right) with the \texttt{wofz} reference.}
 \label{fctError16}
\end{figure*}

The $n=16$ rational approximation is compared to the \texttt{wofz} reference code in \qufig{fctError16}.
Variation of $\delta$ indicates that for $\delta=1.3118$ the approximation is optimal w.r.t.\ overall accuracy, i.e.\ the relative error is $\le 7.86 \cdot 10^{-5}$;
for $y > 10^{-6}$ the error is less than about $10^{-5}$ for all $x,y$ and the largest error is observed at $x \approx 4.8$ and $y=10^{-8}$.
Interestingly, the imaginary part $L(x,y)$ is slightly less accurate compared to the $n=12$ approximation.
For $\delta \approx 1.33$ (and very small $y$) the location of the maximum error moves into the line wings.
For $\delta=1.35$ the maximum of the relative error is slightly larger, about $2\cdot 10^{-4}$, but this maximum is achieved for large $x$ and very small $y$ where the function values are already tiny
($K(5,10^{-6}) \approx 10^{-8}$).
For the imaginary part the maximum relative deviation is $2.6 \cdot 10^{-6}$. 

\begin{figure*}
 \includegraphics[width=\textwidth]{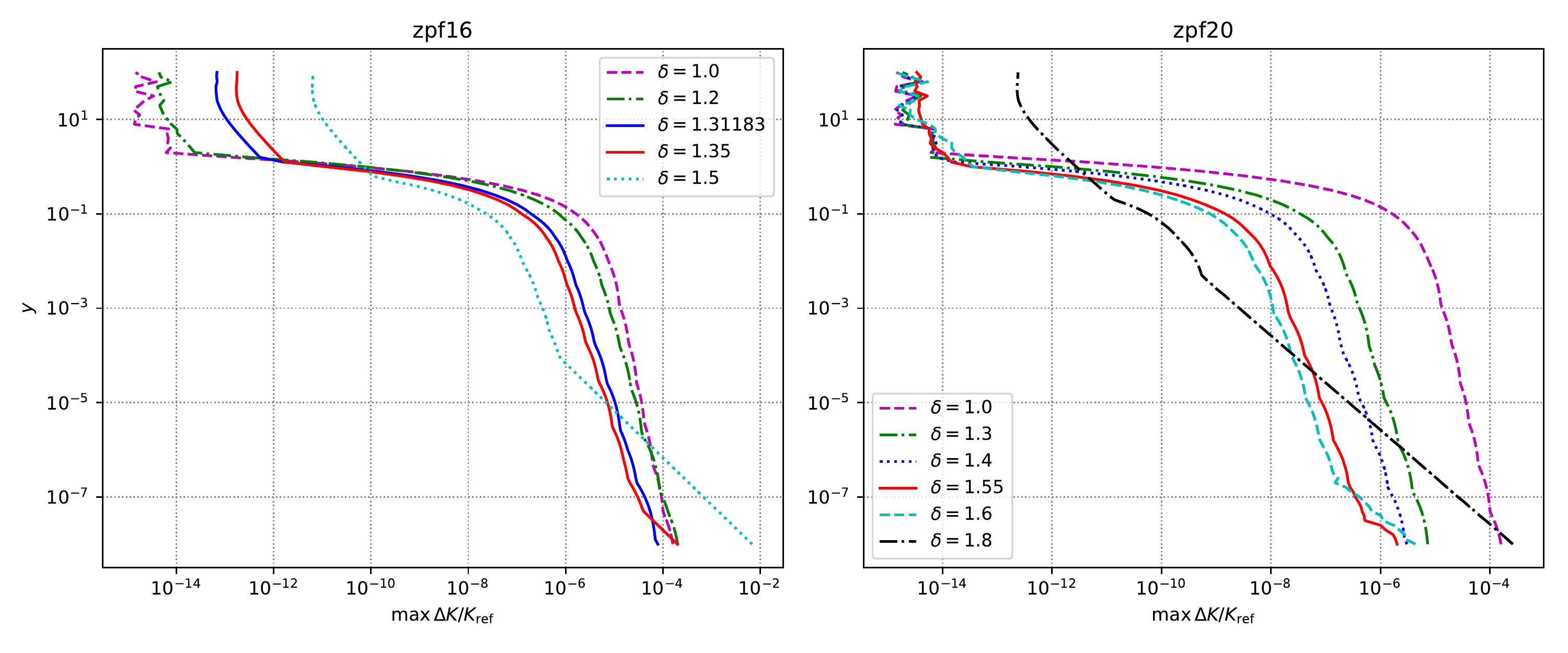}
 \caption{The maximum relative error of the \citeauthor{Humlicek79} $n=16$ and $n=20$ approximations for various $\delta$ compared to the \texttt{wofz} reference.}
 \label{fctErrorY}
\end{figure*}

In \qufig{fctErrorY}a we compare the maximum relative error $\max_x \epsilon(x,y)$ as a function of $y$.
For $y>10^{-4}$ the approximation \eqref{cpf12a} or \eqref{RatFunComplex} is accurate to four significant digits for all $\delta$ considered.

\begin{figure*}
 \includegraphics[width=\textwidth]{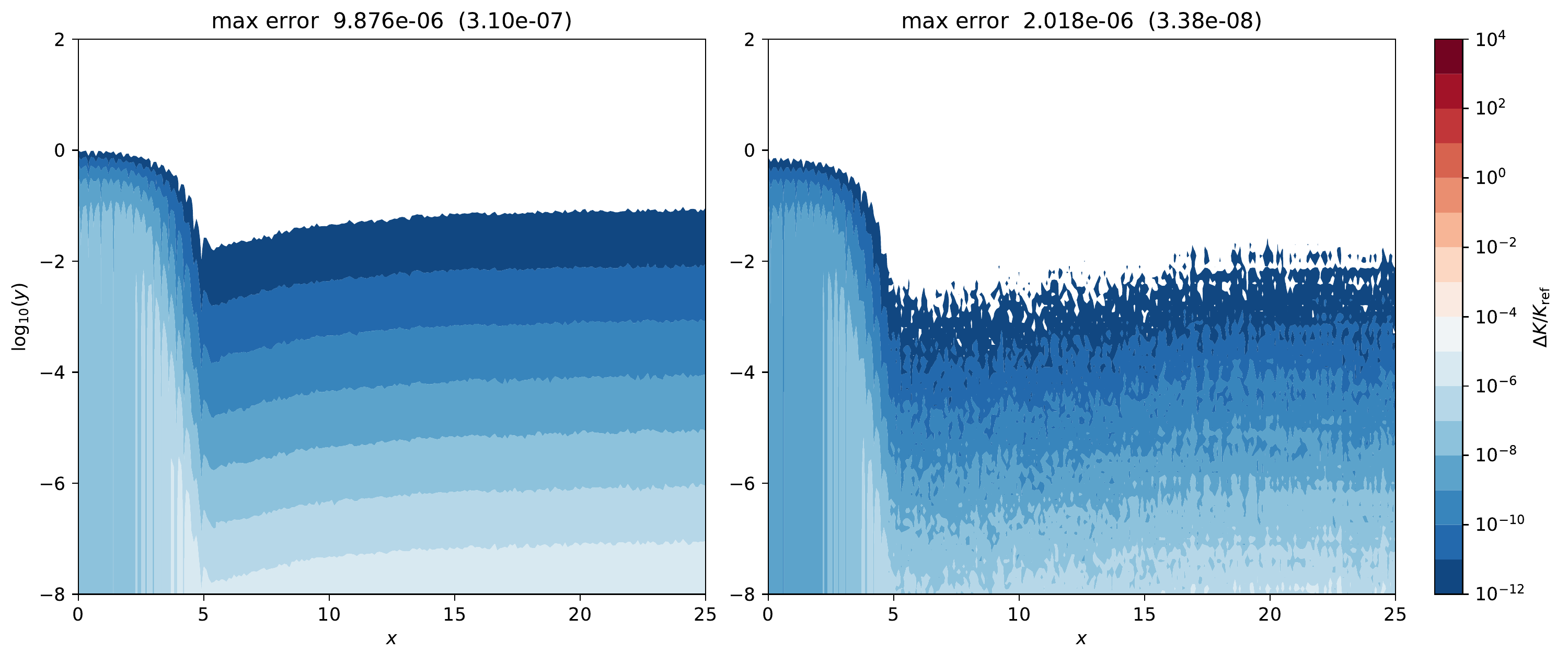}
 \caption{Comparison of \citeauthor{Humlicek79} $n=18$ ($\delta=1.45$, left) and $n=20$ ($\delta=1.55$, right) approximation with the \texttt{wofz} reference.}
 \label{fctError20}
\end{figure*}

Increasing the number of terms in the rational approximation \eqref{cpf12a} further improves the accuracy.
\qufig{fctErrorY}b shows that six significant digits can be achieved with $n=20$ and $\delta \approx 1.55$. 
Contour plots of the relative error of the $n=18$ and $n=20$ approximations with optimal $\delta$ are shown in \qufig{fctError20}.

For a further appraisal of the generalized \citet{Humlicek79} approximation \eqref{cpf12a} we evaluate the Voigt function (i.e.\ $\text{Re}(w))$ for selected values of $z = x + \I y$ and compare these with the results of \citet{Lether91a} who use a corrected midpoint quadrature rule.
The values of $K(x,y)$ compiled in their Table 2 ``are believed to be correct to twenty-five significant digits'' and have also been cross-checked by \citet[][his table 2]{Zaghloul07} and \citet[][table 1]{Boyer14}.
\qutab{tab:acc} shows that for $n=16$ and $n=20$ the expansion \eqref{cpf12a} is correct to at least five significant digits for all $x,y$ pairs except for $x=y=10^{-3}$.
Note that for $y=10^{-20}$ (first data row) the Voigt function is essentially a Gaussian, and the correct value $\exp(-1.0) = 0.36787\;94411\;71442$ is reproduced with six, eight, and nine digits for $n=16$, $n=20$, and $n=24$, respectively.
In the last three rows with moderate $x$ and $y$ the \texttt{cpf} values are correct to ten digits.

\begin{table*}
 \caption{Comparison of Voigt function values for the $n=16$, $n=20$, and $n=24$ approximation \eqref{cpf12a} with the precise values of \citet[][their table 2 (given with 25 digits)]{Lether91a}.}
 \label{tab:acc}
 \begin{tabular}{llllll}
  \hline
  $x$      & $y$        & Lether\,\&\,Wenston          & $n=16 ~(\delta=1.35) $       & $n=20 ~(\delta=1.55)$         & $n=24 ~(\delta=1.4)$ \\
  \hline
  1.0      & $10^{-20}$ & 0.36787 94411 71442          & 0.36787 93403 59605          & 0.36787 94396 18318           & 0.36787 94414 03711          \\
   10.     & $10^{-4}$  & 0.57287 17561 64533 \ten{-6} & 0.57287 17507 11831 \ten{-6} & 0.57287 17561 87614 \ten{-6}  & 0.57287 17561 66040 \ten{-6} \\ 
 $10^{-3}$ & $10^{-3}$  & 0.99887 16233 35411          & 0.99885 13083 62977          & 0.99886 16184 44897           & 0.99887 16170 22162          \\
   0.0     & 0.25       & 0.77034 65477 30996          & 0.77034 65303 12796          & 0.77034 65475 77724           & 0.77034 65475 71460          \\ 
   1.0     & 0.50       & 0.35490 03328 67577          & 0.35490 03328 53826          & 0.35490 03328 64360           & 0.35490 03328 66028          \\ 
   5.0     & 5.0        & 0.56965 43988 71769 \ten{-6} & 0.56965 43988 81711 \ten{-1} & 0.56965 43988 81771 \ten{-1}  & 0.56965 43988 81769 \ten{-1} \\ 
   1.0     & 10.0       & 0.55598 31964 10553 \ten{-6} & 0.55598 31964 10505 \ten{-1} & 0.55598 31964 10555 \ten{-1}  & 0.55598 31964 10553 \ten{-1} \\
  \hline
 \end{tabular}
\end{table*}

\subsection{Preliminary Speed Tests with IPython}
\label{ssec:ipy}

To get a first impression about the performance of the various approximations, we have used IPython's builtin ``magic'' function \texttt{\%timeit}.
A typical ``experiment'' looks like
\begin{verbatim}
In [1]: from cpfX import zpf16p
In [2]: x=numpy.linspace(0.,100.,10001);  y=0.001;
        z=x+1j*y
In [3]: %timeit zpf16p(z)
1000 loops, best of 3: 1.04 ms per loop
\end{verbatim}

\autoref{timeit} summarizes the results of these experiments.
Note that the grid point spacing $\delta x = 0.1$ (first three columns) is more than adequate to sample the line center region as the half width \eqref{xHalf} is approximately $x_{1/2} \approx 1$ for small $y$.  

For the Python translation of the original Fortran \texttt{cpf12} code two versions have been implemented:
A scalar function with an \texttt{if} statement that is transformed in an array function using NumPy's \texttt{vectorize} function class
and an array function exploiting a mask and NumPy's \texttt{where}.
The first two rows of the table clearly indicate that the vectorized version is significantly slower.

The polynomials in \eqref{RatFunComplex} can be implemented either with the Horner scheme coded by hand (\texttt{zpf12h}, \texttt{zpf16h}) or exploiting NumPy's \texttt{poly1d} class (\texttt{zpf12p}, \texttt{zpf16p}), and the \texttt{\%timeit} tests demonstrate that the hand-coded Horner scheme is slightly faster for all cases.
It should also be noted that implementing the denominator as a polynomial of $z^2$ (the odd coefficients are zero) is more efficient than the polynomial of $z$.

For comparison we also report times for several \citet{Weideman94} approximations.
His $N$-term approximation requires $N+4$ multiplications plus one division, so the computational effort for \texttt{weideman32} is similar to \texttt{zpf16}.
However, as mentioned above, the $16$-degree denominator polynomial of \texttt{zpf16} can be implemented more efficiently because of the vanishing odd coefficients,
i.e.\ with 24 add-multiplies for \texttt{zpf16} similar to \texttt{weideman24}.

The \texttt{wofz} implementation documented in the very last row appears to be faster than all rational approximations.
Note, however, that this special function imported from SciPy is actually a C code linked to Python in contrast to our pure Numeric Python codes.

Note that the execution time is approximately proportional to the number of function evaluations only for the \texttt{cpf12v} case (first data row) that is by far the slowest implementation.
Furthermore, times in the fourth column ($n_x=10\,000$) are roughly ten times larger compared to the third column ($n_x=1000$).
The deviations from strict proportionality might indicate that \texttt{\%timeit} estimates also include overhead for function call etc.

\begin{table*}
 \caption{Execution time measured by the \texttt{\% timeit} function in the IPython interpreter. For all columns $y=0.001$ has been used.
 $n_x$ is the number of $x$ grid points and function evaluations.
 The ``zpf'' indicates implementations of the optimized rational approximation \eqref{RatFunComplex}.
 The tests have been performed on a laptop with an Intel x86\_64  CPU running at 2.7\,GHz and cache size 3072\,KB.}  
 \label{timeit}
 \begin{tabular}{lrrrr}
  \hline
             & $0 \le x \le 25$ & $0 \le x \le 50$ & $0 \le x \le 100$ & $0 \le x \le 100$ \\
~~~~~~~~~~~ $n_x$   & 251              & 501              & 1001              & 10001             \\
  \hline
  cpf12v     & 12.1 ms          & 24.8 ms          &  49.4 ms    &  492  ms    \\   
  cpf12w     & 877 $\mu$s       & 1.01 ms          &  1.17 ms    &  4.39 ms    \\[1ex]   
  zpf12h     & 64.2 $\mu$s      & 84 $\mu$s        & 119 $\mu$s  &  829 $\mu$s \\   
  zpf12p     & 116 $\mu$s       & 145 $\mu$s       & 185 $\mu$s  &  1.01 ms    \\[1ex]   
  zpf16h     & 90.1 $\mu$s      & 110 $\mu$s       & 147 $\mu$s  &  904 $\mu$s \\        
  zpf16p     & 117  $\mu$s      & 146 $\mu$s       & 185 $\mu$s  &  1.04 ms    \\[1ex]   
  weideman16 & 78.5 $\mu$s      & 100 $\mu$s       & 134 $\mu$s  &  776  $\mu$s \\
  weideman24 & 105 $\mu$s       & 138 $\mu$s       & 172 $\mu$s  &  1.04 ms    \\
  weideman32 & 132 $\mu$s       & 162 $\mu$s       & 216 $\mu$s  &  1.31 ms    \\[1ex]
  wofz       & 52.6 $\mu$s      & 66.8 $\mu$s      & 91.7 $\mu$s &  886 $\mu$s \\
  \hline
 \end{tabular}
\end{table*}

\subsection{Preformance of lbl cross section modeling}
\label{ssec:lbl}

For realistic benchmarks we test the generalized \citeauthor{Humlicek79} approximation \eqref{RatFunComplex} in an atmospheric radiative transfer lbl modeling context.
Radiative transfer in Earth's atmosphere is clearly important in the geosciences, esp.\ for climate modeling and remote sensing, but is also relevant for astronomy, i.e.\ for corrections of telluric absorption and emission in ground-based observations \citep[e.g.][]{Seifahrt10,Bertaux14,Smette15}
or radiative transfer modeling of Earth-like exoplanets \citep[e.g.][]{DesMarais02,Robinson18}.
To make comparison with our previous studies \citep{Schreier08,Schreier11v} easier, we compute high resolution molecular absorption cross sections
for the SubMillimeter Radiometer (SMR) aboard the Swedish small satellite ODIN \citep{Murtagh02,Nordh03}, an aeronomy and astronomy mission launched in 2001
(the astronomy mission was successfully concluded in spring 2007, see also \citet{Hjalmarson03}). 

The cross sections
\begin{equation}  \label{absXS}
 k(\nu,p,T) ~=~ \sum\limits_l S_l(T) \: g_\text{V}\Bigl(\nu-\hat\nu_l, \gamma_l^\text{(L)}(p,T), \gamma_l^\text{(G)}(T) \Bigr) ~.
\end{equation}
(with line position $\hat\nu_l$ and line strength $S_l$) are computed in the $16 \,\text{--}\,17 \rm\, cm^{-1}$ interval (480 -- 510\,GHz) for thirteen pressure-temperature pairs (corresponding to the $0 \,\text{--}\, 120\rm\,km$ altitude range with $\delta z = 10 \rm\,km$ steps).
Because of the importance of line wing contributions all lines in the extended $6 \,\text{--}\,27 \rm\, cm^{-1}$ interval are considered:
For nitric acid (\chem{HNO_3}) HITRAN\,00 \citep{Rothman03etal} lists 2376 lines (HITRAN\,16 \citep{Gordon17} has about 250\,000 lines), and there are 21\,565 lines of ozone (\chem{O_3}) in
all versions of HITRAN since 1996.
The mean Lorentz to Doppler (Gauss) width ratio $y$ decreases from $7 \cdot 10^3$ (at bottom-of-atmosphere, BoA) to $1.4 \cdot 10^{-4}$ (at top-of-atmosphere, ToA) for \chem{HNO_3};
the Doppler width is slightly larger for \chem{O_3} because of the smaller mass, hence $y$ is somewhat smaller.

\begin{figure*}
 \includegraphics[width=\textwidth]{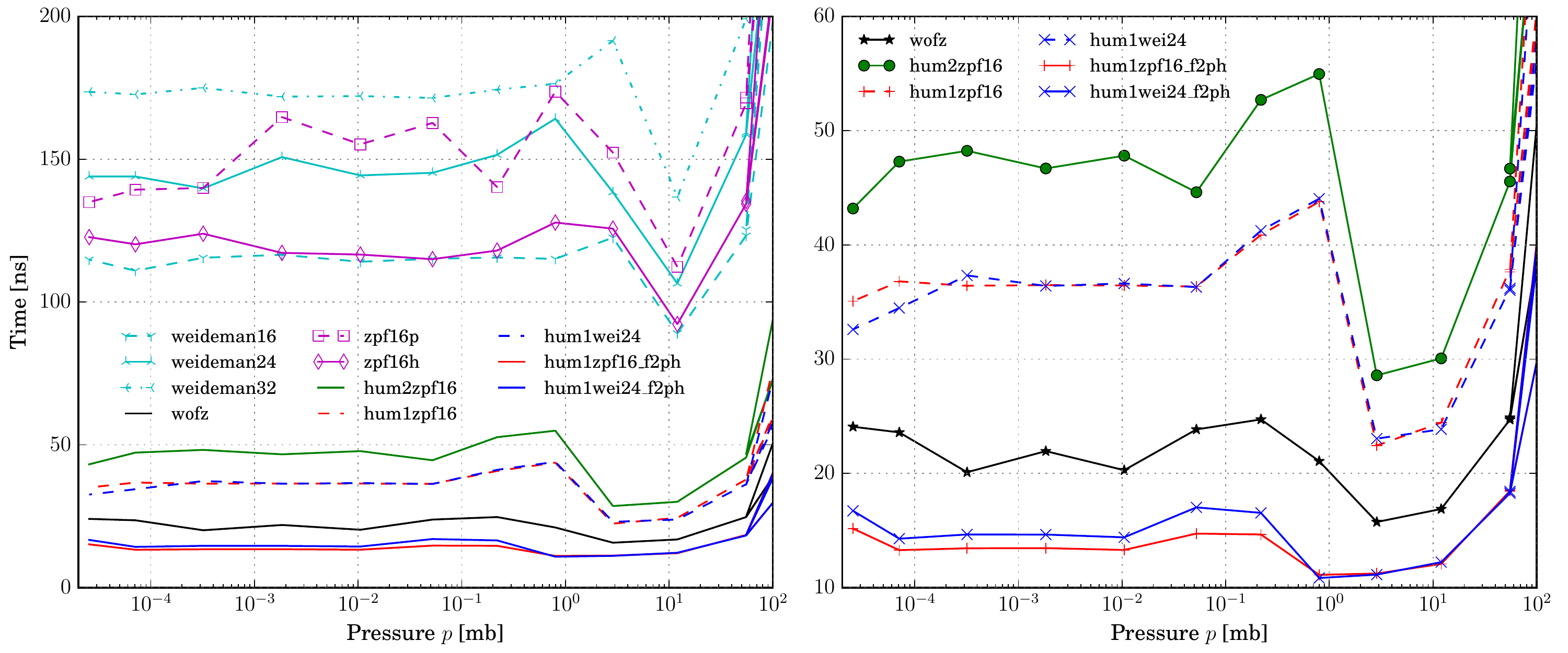}
 \caption{Execution time (ns) per function evaluation.
          The tests have been performed on an idle node of a Linux cluster with an Intel XEON CPU at 2.36\,GHz.}
 \label{time_HNO3}
\end{figure*}

\subsubsection{\chem{HNO_3} cross sections --- Python}
\label{sssec:py}

The \chem{HNO_3} cross sections are modeled using the \texttt{lbl2xs.py} function script of \texttt{Py4CAtS} --- Python for Computational Atmospheric Spectroscopy (available at \url{https://atmos.eoc.dlr.de/tools/Py4CAtS/}) using a ``brute force'' approach, i.e.\ without further approximations in the line wings (e.g.\ coarse grid).
Note that the spectral resolution is depending on line width (essentially pressure, hence altitude), i.e.\ the grid point spacing $\delta x$ is set to a fraction of the mean half width,
typically $\delta x = \bar x_{1/2}/4$ or $\delta\nu = \bar\gamma_\text{V}/4$.
The number of wavenumber grid points increases from less than one hundred at BoA to almost half a million at an altitude of $90\rm\,km$.  

The execution times for a single function value (i.e.\ the total elapsed CPU time measured with the \texttt{clock} function of Python's \texttt{time} module divided by the number of lines and the number of $x$ grid points) shown in \qufig{time_HNO3} essentially confirm the IPython \texttt{\%timeit} tests reported in the previous subsection.
Note that pre- and postprocessing steps (e.g.\ reading the line data) have not been considered for the timing.

Both versions of the \texttt{cpf12} code are significantly slower than all other functions:  the \texttt{cpf12w} code needs about 600\,ns for $p \le 10\rm\,mb$ and is not shown.
Except for large pressures the various Weideman functions and the two implementations of the approximation \eqref{RatFunComplex} with $n=16$ require about $100$ to $175\rm\,ns$,
and the use of an asymptotic approximation for large $x$ is tempting.
The performance of the combination \eqref{hum1cpf16} as well as the \citet{Humlicek82}--\citet{Weideman94} combination (with $N=24$) are almost identical,
and somewhat worse to the \texttt{wofz} C function linked to Python.
The combination of the $R_{3,4}$ rational approximation of \citet{Humlicek82} with the \citeauthor{Weideman94} $N=32$ approximation (with a cut at $s=8$, cf.\ \citet{Schreier17}) has a relative error better than $2 \cdot 10^{-6}$ and is slightly slower.

The good performance of the \texttt{wofz} C function reported in the previous subsection suggests that implementation of compute intensive code segments in a compiled language might be advantageous.
Accordingly we have used the \texttt{f2py} Fortran-Python interface generator \citep{Peterson09} to make calls to Fortran subroutines possible.
Both combinations of the \citet{Humlicek82} $R_{1,2}$ asymptotic approximation \eqref{w4asymp} with the \citet{Humlicek79} \texttt{zpf16} approximation \eqref{RatFunComplex} or with the 24-term Weideman approximation (denoted \texttt{hum1zpf16\_f2py} and \texttt{hum1wei24\_f2py}) are more than a factor two faster than the corresponding pure Python implementation and clearly superior with respect to computational efficiency.

Compared to the tests reported in \citet[][Fig.\ 11]{Schreier11v} the \citeauthor{Humlicek82}--\citeauthor{Weideman94} as well as the pure Weideman functions are about a factor 10 faster now.
Note that the minimum times seen here are comparable to the Fortran execution times shown in Fig.\ 10 of \citet{Schreier11v}.
This considerable speed-up is presumably a combination of better hardware and better software (improved performance of NumPy).

\subsubsection{\chem{O_3} cross sections --- Fortran}
\label{sssec:for}

The \citet{Humlicek79} $n=16$ approximation \eqref{RatFunComplex} along with several other complex error function algorithms has also be implemented in Fortran\,90.
The execution time for the lbl modeling of \chem{O_3} cross sections is measured by means of the \texttt{cpu\_time} intrinsic function.

In contrast to the Python tests the runtime appears to be fairly constant for pressures smaller than about $10\rm\,mb$ where the number of wavenumber grid points is already large ($n_x>10^4$).
Hence we only report the average run time in \autoref{time_O3_fortran}.
\texttt{cpf12} roughly corresponds to the original code documented in the \citet{Humlicek79} paper and Fortran can execute this code significantly faster compared to NumPy.
The optimized implementation \eqref{RatFunComplex} of the 16-term approximation \texttt{zpf16} is executed in Fortran with roughly double speed.
The combinations of the \texttt{zpf16} approximation \eqref{RatFunComplex} or Weideman's approximation with the \citet{Humlicek82} $R_{12}$ or $R_{34}$ approximation for large $|x|$ have approximately identical run times.
Comparison with \autoref{time_HNO3} reveals that the speed of the Fortran and NumPy--f2py codes is roughly similar, i.e.\ $15\rm\,ns$ per function value.
The two versions of the GNU gfortran compiler produce code with approximately the same performance, whereas the NAG compiled code is somewhat slower (similar to \citet[][Fig.\ 9]{Schreier11v}).

There appears to be no pure Fortran array implementation of \texttt{wofz}.
 The \texttt{libcerf} numeric library (\url{http://apps.jcns.fz-juelich.de/libcerf}) implementation of the Faddeeva package also provides Fortran bindings, but only for scalar arguments.
 However, the execution time of the \texttt{hum1zpf16} and \texttt{hum1wei24} pure Fortran subroutines and the functions wrapped to Python via f2py are almost identical,
 so we can expect that a Fortran array version of \texttt{wofz} would perform similar to the SciPy-C version that was about a factor 1.5 slower compared to \texttt{hum1zpf16} and \texttt{hum1wei24}
 (compare \autoref{time_HNO3} and \autoref{time_O3_fortran}).

\begin{table}
 \caption{Execution time [ns] per function evaluation for the Fortran\,90 implementation.
          GNU Fortran or NAG compiler with optimization flag \texttt{-O3} (the numbers indicate the version). Linux cluster node as in \qufig{time_HNO3}.}
 \label{time_O3_fortran}
 \begin{tabular}{lrrr}
  \hline
             & gfortran 4.8 & gfortran 6.2 & nagfor 5.3 \\
  \hline
  cpf12      & 97           & 102.6        & 125        \\
  zpf16      & 64.6         & 61.2         & 66.24      \\
  hum1zpf16  & 14.52        & 14.61        & 16.98      \\
  hum1wei24  & 14.55        & 14.49        & 16.39      \\
  hum2wei32  & 14.75        & 14.86        & 18.15      \\
  weideman16 & 70.34        &              &            \\
  weideman32 & 131.1        &              &            \\
  \hline
 \end{tabular}
\end{table}

\section{Summary and Conclusions}
\label{sec:conclusions}

Generalizations of the \citet{Humlicek79} \texttt{cpf12} rational approximation (with 12 terms) for the Voigt and complex error function have been presented and optimized implementations have been developed.
The rational approximation with a degree 16 denominator polynomial is accurate to at least five significant digits except for very small $y$, and somewhat larger deviations show up in the wings where the function value is already extremely small.
For $n \ge 16$ the correction term needed to compensate for the problems of the $n=12$ approximation \eqref{cpf12a} for large $|x|$ and small $y$ is not required anymore,
i.e.\ a single rational approximation can be used to evaluate the complex error function over almost the entire complex plane.

The sum of $n$ fractions of the original approximation of \citet{Humlicek79} are computationally inefficient, but can be readily transformed into a single fraction with an $n-1$ degree numerator polynomial and an $n$ degree
denominator polynomial.
The performance of the original and generalized \citeauthor{Humlicek79} algorithms implemented as Numeric Python functions has been tested and compared to other rational approximations.
For a more realistic scenario molecular absorption cross sections have been computed line-by-line using Python and Fortran implementations.
The combination of the $n=16$ approximation with the asymptotic approximation also developed by \citet{Humlicek82} further improved the performance.
The combination of the \citet{Humlicek79,Humlicek82} methods as well as the \citeauthor{Humlicek79}--\citeauthor{Weideman94} combination need about $15 \rm\,ns$ for a single function value both with Fortran and Python,
in strong contrast to our findings seven years ago \citep{Schreier11v} where Fortran was about a factor ten faster.
This impressive speed of the Python implementation on a par with the Fortran code explains the increasing popularity of the ``Python-based ecosystem'' (\url{www.scipy.org}) and packages such as Astropy \citep{astropy13etal}. 

In conclusion, the combinations of the \citet{Humlicek79} or \citet{Weideman94} approximation with the \citet{Humlicek82} asymptotic approximation can be recommended for
both accurate (five or more digits) and efficient Voigt and complex error function evaluations.
For high performance with Python, a Fortran (or C) implementation linked to Python is advantageous.
When speed is not an issue, the generalization of the original \texttt{cpf12} code to 16 or more terms can be used for all $x,y$ without any conditional branches.


\section*{Supplementary material}
Two Python source files are provided as supplementary material with the online version of this paper. 
Additionally we provide the Fortran code used for the tests of subsection \ref{sssec:for} including the \citet{Humlicek79,Humlicek82} combination \eqref{hum1cpf16}.
\begin{description}
 \item{\tt cpfX.py} Various implementations of the \citet{Humlicek79} algorithm:
  \begin{itemize}
   \item Python/NumPy implementations of the original Fortran 77 source code (with two versions to combine the two regions);
   \item  A generalization of the region I rational approximation to arbitrary (even) number of terms;
   \item An optimized implementation of the region I approximation for 16 terms using a single fraction; 
   \item The combination \eqref{hum1cpf16} for NumPy.
  \end{itemize}
 \item{\tt fig124.py} Execution of this script inside the IPython interpreter should produce the Figures 1, 2, and 4 of the manuscript.
 \item{\tt Test\_voigt\_speed.f90} The Fortran\,90 program including subroutines of various complex error function algorithms.

\end{description}


\section*{Acknowledgements}
Financial support by the DFG project SCHR 1125/3-1 is greatly appreciated.
I would also like to thank Thomas Trautmann for critical reading of the manuscript.
The transformation of the sum of quotients \eqref{cpf12a} to the optimized form \eqref{RatFunComplex} has been computed with \sc{SymPy} (\url{http://www.sympy.org/}).


\begin{thebibliography}{}
\makeatletter
\relax
\def\mn@urlcharsother{\let\do\@makeother \do\$\do\&\do\#\do\^\do\_\do\%\do\~}
\def\mn@doi{\begingroup\mn@urlcharsother \@ifnextchar [ {\mn@doi@}
  {\mn@doi@[]}}
\def\mn@doi@[#1]#2{\def\@tempa{#1}\ifx\@tempa\@empty \href
  {http://dx.doi.org/#2} {doi:#2}\else \href {http://dx.doi.org/#2} {#1}\fi
  \endgroup}
\def\mn@eprint#1#2{\mn@eprint@#1:#2::\@nil}
\def\mn@eprint@arXiv#1{\href {http://arxiv.org/abs/#1} {{\tt arXiv:#1}}}
\def\mn@eprint@dblp#1{\href {http://dblp.uni-trier.de/rec/bibtex/#1.xml}
  {dblp:#1}}
\def\mn@eprint@#1:#2:#3:#4\@nil{\def\@tempa {#1}\def\@tempb {#2}\def\@tempc
  {#3}\ifx \@tempc \@empty \let \@tempc \@tempb \let \@tempb \@tempa \fi \ifx
  \@tempb \@empty \def\@tempb {arXiv}\fi \@ifundefined
  {mn@eprint@\@tempb}{\@tempb:\@tempc}{\expandafter \expandafter \csname
  mn@eprint@\@tempb\endcsname \expandafter{\@tempc}}}

\bibitem[\protect\citeauthoryear{Abramowitz \& Stegun}{Abramowitz \&
  Stegun}{1964}]{AbSt64}
Abramowitz M.,  Stegun I.,  1964, Handbook of Mathematical Functions.
National Bureau of Standards, AMS55, New York

\bibitem[\protect\citeauthoryear{Armstrong}{Armstrong}{1967}]{Armstrong67}
Armstrong B.,  1967, \mn@doi [JQSRT] {10.1016/0022-4073(67)90057-X}, 7, 61

\bibitem[\protect\citeauthoryear{{Astropy Collaboration}, Robitaille, Tollerud
  et~al.}{{Astropy Collaboration}}{2013}]{astropy13etal}
{Astropy Collaboration} T.,  Robitaille T.,  Tollerud E.,   et~al., 2013,
  \mn@doi [A\&A] {10.1051/0004-6361/201322068}, 558, A33

\bibitem[\protect\citeauthoryear{Bailey \& Kedziora-Chudczer}{Bailey \&
  Kedziora-Chudczer}{2012}]{Bailey12}
Bailey J.,  Kedziora-Chudczer L.,  2012, \mn@doi [MNRAS]
  {10.1111/j.1365-2966.2011.19845.x}, 419, 1913

\bibitem[\protect\citeauthoryear{Bertaux, Lallement, Ferron, Boonne  \&
  Bodichon}{Bertaux et~al.}{2014}]{Bertaux14}
Bertaux L.,  Lallement R.,  Ferron S.,  Boonne C.,   Bodichon R.,  2014,
  \mn@doi [A\&A] {10.1051/0004-6361/201322383}, 564, A46

\bibitem[\protect\citeauthoryear{Boyer \& Lynas-Gray}{Boyer \&
  Lynas-Gray}{2014}]{Boyer14}
Boyer W.,  Lynas-Gray A.,  2014, \mn@doi [MNRAS] {10.1093/mnras/stu1606}, 444,
  2555

\bibitem[\protect\citeauthoryear{{Des Marais} et~al.,}{{Des Marais}
  et~al.}{2002}]{DesMarais02}
{Des Marais} D.,  et~al., 2002, \mn@doi [Astrobiology]
  {10.1089/15311070260192246}, 2, 153

\bibitem[\protect\citeauthoryear{Goedecker \& Hoisie}{Goedecker \&
  Hoisie}{2001}]{Goedecker01}
Goedecker S.,  Hoisie A.,  2001, Performance Optimization of Numerically
  Intensive Codes.
SIAM, Philadelphia, PA

\bibitem[\protect\citeauthoryear{Gordon et~al.,}{Gordon
  et~al.}{2017}]{Gordon17}
Gordon I.,  et~al., 2017, \mn@doi [JQSRT] {10.1016/j.jqsrt.2017.06.038}, 203, 3

\bibitem[\protect\citeauthoryear{Grimm \& Heng}{Grimm \& Heng}{2015}]{Grimm15}
Grimm S.~L.,  Heng K.,  2015, \mn@doi [ApJ] {10.1088/0004-637X/808/2/182}, 808,
  182

\bibitem[\protect\citeauthoryear{Heng}{Heng}{2017}]{Heng17b}
Heng K.,  2017, Exoplanetary Atmospheres --- Theoretical Concepts and
  Foundations.
Princeton University Press

\bibitem[\protect\citeauthoryear{Hjalmarson et~al.,}{Hjalmarson
  et~al.}{2003}]{Hjalmarson03}
Hjalmarson A.,  et~al., 2003, \mn@doi [A\&A] {10.1051/0004-6361:20030337}, 402,
  L39

\bibitem[\protect\citeauthoryear{Hui, Armstrong  \& Wray}{Hui
  et~al.}{1978}]{Hui78}
Hui A.,  Armstrong B.,   Wray A.,  1978, \mn@doi [JQSRT]
  {10.1016/0022-4073(78)90019-5}, 19, 509

\bibitem[\protect\citeauthoryear{Huml\'\i\v{c}ek}{Huml\'\i\v{c}ek}{1979}]{Humlicek79}
Huml\'\i\v{c}ek J.,  1979, \mn@doi [JQSRT] {10.1016/0022-4073(79)90062-1}, 21,
  309

\bibitem[\protect\citeauthoryear{Huml\'\i\v{c}ek}{Huml\'\i\v{c}ek}{1982}]{Humlicek82}
Huml\'\i\v{c}ek J.,  1982, \mn@doi [JQSRT] {10.1016/0022-4073(82)90078-4}, 27,
  437

\bibitem[\protect\citeauthoryear{Imai, Suzuki  \& Takahashi}{Imai
  et~al.}{2010}]{Imai10}
Imai K.,  Suzuki M.,   Takahashi C.,  2010, \mn@doi [Adv.\ Space Res.]
  {10.1016/j.asr.2009.11.005}, 45, 669

\bibitem[\protect\citeauthoryear{Jacquinet-Husson et~al.,}{Jacquinet-Husson
  et~al.}{2016}]{JacquinetHusson16}
Jacquinet-Husson N.,  et~al., 2016, \mn@doi [J.\ Mol.\ Spectrosc.]
  {10.1016/j.jms.2016.06.007}, 327, 31

\bibitem[\protect\citeauthoryear{Karp}{Karp}{1978}]{Karp78}
Karp A.,  1978, \mn@doi [JQSRT] {10.1016/0022-4073(78)90106-1}, 20, 379

\bibitem[\protect\citeauthoryear{Kochanov}{Kochanov}{2011}]{Kochanov11v}
Kochanov V.,  2011, \mn@doi [Atmospheric and Oceanic Optics]
  {10.1134/S1024856011050071}, 24, 432

\bibitem[\protect\citeauthoryear{Kuntz}{Kuntz}{1997}]{Kuntz97}
Kuntz M.,  1997, \mn@doi [JQSRT] {10.1016/S0022-4073(96)00162-8}, 57, 819

\bibitem[\protect\citeauthoryear{Lether \& Wenston}{Lether \&
  Wenston}{1991}]{Lether91a}
Lether F.,  Wenston P.,  1991, \mn@doi [J.\ Comp.\ Appl.\ Math.]
  {10.1016/0377-0427(91)90149-E}, 34, 75

\bibitem[\protect\citeauthoryear{Lynas-Gray}{Lynas-Gray}{1993}]{Lynas-Gray93}
Lynas-Gray A.,  1993, \mn@doi [Comp.\ Phys.\ Comm.]
  {10.1016/0010-4655(93)90171-8}, 75, 135

\bibitem[\protect\citeauthoryear{Molin}{Molin}{2011}]{Molin11}
Molin P.,  2011, Multi-precision computation of the complex error function,
  Preprint at \url{https://hal.archives-ouvertes.fr/hal-00580855}

\bibitem[\protect\citeauthoryear{Murtagh et~al.,}{Murtagh
  et~al.}{2002}]{Murtagh02}
Murtagh D.,  et~al., 2002, \mn@doi [Can.\ J.\ Phys.] {10.1139/p01-157}, 80, 309

\bibitem[\protect\citeauthoryear{Nordh et~al.,}{Nordh et~al.}{2003}]{Nordh03}
Nordh H.~L.,  et~al., 2003, \mn@doi [A\&A] {10.1051/0004-6361:20030334}, 402,
  L21

\bibitem[\protect\citeauthoryear{Olivero \& Longbothum}{Olivero \&
  Longbothum}{1977}]{Olivero77}
Olivero J.,  Longbothum R.,  1977, \mn@doi [JQSRT]
  {10.1016/0022-4073(77)90161-3}, 17, 233

\bibitem[\protect\citeauthoryear{Olver, Lozier, Boisvert  \& Clark}{Olver
  et~al.}{2010}]{NIST-HMF}
Olver F.,  Lozier D.,  Boisvert R.,   Clark C.,  eds, 2010, {NIST} Handbook of
  Mathematical Functions.
Cambridge University Press, New York, NY

\bibitem[\protect\citeauthoryear{Peraiah}{Peraiah}{2002}]{Peraiah02}
Peraiah A.,  2002, An Introduction to Radiative Transfer: Methods and
  Applications in Astrophysics.
Cambridge University Press

\bibitem[\protect\citeauthoryear{Peterson}{Peterson}{2009}]{Peterson09}
Peterson P.,  2009, \mn@doi [Int.\ J. Comp.\ Sci.\ Eng.]
  {10.1504/IJCSE.2009.029165}, 4, 296

\bibitem[\protect\citeauthoryear{Poppe \& Wijers}{Poppe \&
  Wijers}{1990a}]{Poppe90}
Poppe G.,  Wijers C.,  1990a, \mn@doi [ACM TOMS] {10.1145/77626.77629}, 16, 38

\bibitem[\protect\citeauthoryear{Poppe \& Wijers}{Poppe \&
  Wijers}{1990b}]{Poppe90a}
Poppe G.,  Wijers C.,  1990b, \mn@doi [ACM TOMS] {10.1145/77626.77630}, 16, 47

\bibitem[\protect\citeauthoryear{Robinson \& Reinhard}{Robinson \&
  Reinhard}{2018}]{Robinson18}
Robinson T.,  Reinhard C.,  2018, preprint, \href
  {http://adsabs.harvard.edu/abs/2018arXiv180404138R} {} (\mn@eprint {arXiv}
  {1804.04138})

\bibitem[\protect\citeauthoryear{Rothman et~al.}{Rothman
  et~al.}{2003}]{Rothman03etal}
Rothman L.,  et~al., 2003, \mn@doi [JQSRT] {10.1016/S0022-4073(03)00146-8}, 82,
  5

\bibitem[\protect\citeauthoryear{Rothman et~al.,}{Rothman
  et~al.}{2010}]{Rothman10}
Rothman L.,  et~al., 2010, \mn@doi [JQSRT] {10.1016/j.jqsrt.2010.05.001}, 111,
  2139

\bibitem[\protect\citeauthoryear{Ruyten}{Ruyten}{2004}]{Ruyten04}
Ruyten W.,  2004, \mn@doi [JQSRT] {10.1016/j.jqsrt.2003.12.027}, 86, 231

\bibitem[\protect\citeauthoryear{Schreier}{Schreier}{2011}]{Schreier11v}
Schreier F.,  2011, \mn@doi [JQSRT] {10.1016/j.jqsrt.2010.12.010}, 112, 1010

\bibitem[\protect\citeauthoryear{Schreier}{Schreier}{2017}]{Schreier17}
Schreier F.,  2017, \mn@doi [JQSRT] {10.1016/j.jqsrt.2016.08.009}, 187, 44

\bibitem[\protect\citeauthoryear{Schreier \& Kohlert}{Schreier \&
  Kohlert}{2008}]{Schreier08}
Schreier F.,  Kohlert D.,  2008, \mn@doi [Comp.\ Phys.\ Comm.]
  {10.1016/j.cpc.2008.04.012}, 179, 457

\bibitem[\protect\citeauthoryear{Schreier, {Gimeno Garc{\'\i}a}, Hedelt, Hess,
  Mendrok, Vasquez  \& Xu}{Schreier et~al.}{2014}]{Schreier14}
Schreier F.,  {Gimeno Garc{\'\i}a} S.,  Hedelt P.,  Hess M.,  Mendrok J.,
  Vasquez M.,   Xu J.,  2014, \mn@doi [JQSRT] {10.1016/j.jqsrt.2013.11.018},
  137, 29

\bibitem[\protect\citeauthoryear{Seifahrt, K\"aufl, Z\"angl, Bean, Richter  \&
  Siebenmorgen}{Seifahrt et~al.}{2010}]{Seifahrt10}
Seifahrt A.,  K\"aufl H.,  Z\"angl G.,  Bean J.,  Richter M.,   Siebenmorgen
  R.,  2010, \mn@doi [A\&A] {10.1051/0004-6361/200913782}, 524, A11

\bibitem[\protect\citeauthoryear{Shippony \& Read}{Shippony \&
  Read}{1993}]{Shippony93}
Shippony Z.,  Read W.,  1993, \mn@doi [JQSRT] {10.1016/0022-4073(93)90031-C},
  50, 635

\bibitem[\protect\citeauthoryear{Shippony \& Read}{Shippony \&
  Read}{2003}]{Shippony03}
Shippony Z.,  Read W.,  2003, \mn@doi [JQSRT] {10.1016/S0022-4073(02)00169-3},
  78, 255

\bibitem[\protect\citeauthoryear{Smette et~al.,}{Smette
  et~al.}{2015}]{Smette15}
Smette A.,  et~al., 2015, \mn@doi [A\&A] {10.1051/0004-6361/201423932}, 576,
  A77

\bibitem[\protect\citeauthoryear{Tennyson \& Yurchenko}{Tennyson \&
  Yurchenko}{2012}]{Tennyson12}
Tennyson J.,  Yurchenko S.~N.,  2012, \mn@doi [MNRAS]
  {10.1111/j.1365-2966.2012.21440.x}, 425, 21

\bibitem[\protect\citeauthoryear{Tennyson et~al.,}{Tennyson
  et~al.}{2014}]{Tennyson14}
Tennyson J.,  et~al., 2014, \mn@doi [Pure Appl.\ Chem.]
  {10.1515/pac-2014-0208}, 86, 1931

\bibitem[\protect\citeauthoryear{Tepper-Garc\'{\i}a}{Tepper-Garc\'{\i}a}{2006}]{TepperGarcia06}
Tepper-Garc\'{\i}a T.,  2006, \mn@doi [MNRAS]
  {10.1111/j.1365-2966.2006.10450.x}, 369, 2025

\bibitem[\protect\citeauthoryear{Tepper-Garc\'{\i}a}{Tepper-Garc\'{\i}a}{2007}]{TepperGarcia07e}
Tepper-Garc\'{\i}a T.,  2007, \mn@doi [MNRAS]
  {10.1111/j.1365-2966.2007.12186.x}, 382, 1375

\bibitem[\protect\citeauthoryear{Tran, Ngo  \& Hartmann}{Tran
  et~al.}{2013}]{Tran13}
Tran H.,  Ngo N.,   Hartmann J.-M.,  2013, \mn@doi [JQSRT]
  {10.1016/j.jqsrt.2013.06.015}, 129, 199

\bibitem[\protect\citeauthoryear{Ueberhuber}{Ueberhuber}{1997}]{Ueberhuber97}
Ueberhuber C.,  1997, Numerical Computation.
Springer

\bibitem[\protect\citeauthoryear{Varghese \& Hanson}{Varghese \&
  Hanson}{1984}]{Varghese84}
Varghese P.,  Hanson R.,  1984, \mn@doi [Appl.\ Opt.] {10.1364/AO.23.002376},
  23, 2376

\bibitem[\protect\citeauthoryear{Weideman}{Weideman}{1994}]{Weideman94}
Weideman J.,  1994, \mn@doi [SIAM J.\ Num.\ Anal.] {10.1137/0731077}, 31, 1497

\bibitem[\protect\citeauthoryear{Wells}{Wells}{1999}]{Wells99}
Wells R.,  1999, \mn@doi [JQSRT] {10.1016/S0022-4073(97)00231-8}, 62, 29

\bibitem[\protect\citeauthoryear{Zaghloul}{Zaghloul}{2007}]{Zaghloul07}
Zaghloul M.~R.,  2007, \mn@doi [MNRAS] {10.1111/j.1365-2966.2006.11377.x}, 375,
  1043

\bibitem[\protect\citeauthoryear{Zaghloul}{Zaghloul}{2017}]{Zaghloul17}
Zaghloul M.,  2017, \mn@doi [ACM TOMS] {10.1145/3119904}, 44, 22:1

\bibitem[\protect\citeauthoryear{Zaghloul \& Ali}{Zaghloul \&
  Ali}{2011}]{Zaghloul11}
Zaghloul M.,  Ali A.,  2011, \mn@doi [ACM TOMS] {10.1145/2049673.2049679}, 38,
  15:1

\makeatother
\end{thebibliography}
\input humlicek_schreier_mnras.bbl

\bsp	
\label{lastpage}

\end{document}